\RequirePackage{ifpdf}
\ifpdf 
\documentclass[pdftex]{sigma}
\else
\documentclass{sigma}
\fi

\numberwithin{equation}{section}

\def\HH{\mathcal{H}}
\def\K{\mathcal{K}}

\def\n{{\cal N}}

\newcommand{\ga}{\gamma}

\newcommand{\al}{\alpha}

\newcommand{\lga}{\longrightarrow}

\def\barra#1{\not \!#1}

\def\H{\mathbb H}
\def\BC{\mathbb C}
\def\_\BC{\mathbbi C}

\def\SD{\rtimes}

\def\H{\mathbb{H}}
\def\R{\mathbb{R}}
\def\N{\mathbb{N}}
\def\M{\mathbb{M}}
\def\C{\mathbb{C}}
\def\Z {\mathbb{Z}}

\def\ID{\mathbb I}

\def\spt {space-time}

\def\minks {Minkowski spacetime}

\def\dS {de Sitter}

\def\qft {quantum field theory}

\newcommand{\deq}{\stackrel{\mathrm{def}}{=}}

\newcommand\dpx[1]{\frac{\partial}{\partial#1}}
\newcommand\ddpx[1]{\frac{\partial^2}{\partial#1^2}}

\def\bee{\begin{enumerate}}
\def\eee{\end{enumerate}}

\def\le{\langle}
\def\re{\rangle}

\newcommand{\bei}{\begin{itemize}}
\newcommand{\eni}{\end{itemize}}

\newcommand\ben{\begin{enumerate}}
\newcommand\een{\end{enumerate}}

\newcommand{\mat}[1]{\left( \begin{array}{rr} #1 \end{array} \right)}

\newcommand{\<}{\langle}
\renewcommand{\>}{\rangle}
\newcommand{\spa}{{{\rm span}}}
\newcommand{\HC}{{\mathbb{H}}}
\newcommand{\gh}{\mathfrak{h}}

\begin{document}

\allowdisplaybreaks

\renewcommand{\thefootnote}{$\star$}

\renewcommand{\PaperNumber}{011}

\FirstPageHeading

\ShortArticleName{Krein Spaces in de Sitter Quantum Theories}

\ArticleName{Krein Spaces in de Sitter Quantum Theories\footnote{This paper is a
contribution to the Proceedings of the 5-th Microconference
``Analytic and Algebraic Me\-thods~V''. The full collection is
available at
\href{http://www.emis.de/journals/SIGMA/Prague2009.html}{http://www.emis.de/journals/SIGMA/Prague2009.html}}}

\Author{Jean-Pierre GAZEAU~$^\dag$, Petr SIEGL~$^{\dag\ddag}$  and Ahmed YOUSSEF~$^\dag$}

\AuthorNameForHeading{J.-P.~Gazeau, P.~Siegl and A.~Youssef}

\Address{$^\dag$~Astroparticules et Cosmologie (APC, UMR 7164), Universit\'e Paris-Diderot,\\
 \hphantom{$^\dag$}~Boite 7020, 75205 Paris Cedex 13, France}
\EmailD{\href{mailto:gazeau@apc.univ-paris7.fr}{gazeau@apc.univ-paris7.fr}, \href{mailto:psiegl@apc.univ-paris7.fr}{psiegl@apc.univ-paris7.fr}, \href{mailto:youssef@apc.univ-paris7.fr}{youssef@apc.univ-paris7.fr}}

\Address{$^\ddag$~Nuclear Physics Institute of Academy of Sciences of the Czech Republic,\\
\hphantom{$^\ddag$}~250 68 \v Re\v z, Czech Republic}

\ArticleDates{Received October 19, 2009, in f\/inal form January 15, 2010;  Published online January 27, 2010}

\Abstract{Experimental evidences and theoretical motivations lead to  consider the curved space-time relativity based on the de Sitter group $SO_0(1,4)$ or $Sp(2,2)$ as an  appealing substitute to the f\/lat space-time Poincar\'e relativity. Quantum elementary systems are then associated to unitary irreducible representations of that simple Lie group. At the lowest limit of the discrete series  lies a remarkable family of scalar representations involving Krein structures and related undecomposable representation cohomology which deserves to be thoroughly studied in view of quantization of the corresponding carrier f\/ields. The purpose of this note is to present the mathematical material needed to examine the problem and to indicate possible extensions of an exemplary case, namely the so-called de Sitterian massless minimally coupled f\/ield, i.e.\ a scalar f\/ield in de Sitter space-time which does not couple to the Ricci curvature.}


\Classification{81T20; 81R05; 81R20; 22E70; 20C35}

\renewcommand{\thefootnote}{\arabic{footnote}}
\setcounter{footnote}{0}

\section{Introduction}

De Sitter and anti de Sitter space-times  are, with Minkowski space-time, the only maximally symmetric
space-time solutions in general relativity.  Their
respective invariance (in the relativity or kinematical sense) groups
are the ten-parameter de Sitter  $SO_{ 0}(1,4)$ and anti de Sitter $SO_{0}(2,3)$ groups.
Both may be viewed as   deformations of the
proper orthochronous Poincar\'e group $\mathcal{P}^{1,3}\SD\, SO_{
0}(1,3)$, the kinematical group of Minkowski space-time.

The de Sitter (resp.\ anti de Sitter)  \spt s are   solutions
to the vacuum Einstein's equations with
positive (resp.\ negative) cosmological constant $\Lambda$. This
constant is linked to the (constant) Ricci  curvature $4 \Lambda$ of these
\spt s. The  corresponding fundamental length is given by
\begin{gather}
\label{radius}
R = \sqrt{\frac{3}{\vert \Lambda \vert}} = c H^{-1},
\end{gather}
where $H$ is the Hubble constant\footnote{Throughout this text, for convenience,  we will mostly work in units $c = 1 = \hbar$, for which $R = H^{-1}$, while restoring physical units when is necessary.}.

Serious reasons back up any interest in studying Physics in
such constant curvature spacetimes with maximal symmetry. The f\/irst
one is the simplicity of their geometry, which makes us consider them as
an excellent laboratory model in view of studying Physics in more
elaborate universes, more precisely with the purpose to set up a quantum f\/ield theory as much rigorous as possible \cite{Isham,Fulling1989,Wald94}.
In this paper we are only interested in the de Sitter space-time. Indeed, since the beginning of
the eighties,  the de Sitter space, specially the spatially f\/lat version of it, has been
playing a much popular role in inf\/lationary cosmological scenarii
 where it is assumed that the cosmic dynamics was
dominated by a term acting like a cosmological constant.
More recently, observations on far high redshift
supernovae, on galaxy clusters, and on
cosmic mic\-rowave background radiation  suggested an
accelerating universe. Again, this can be explained with such a
term. For  updated reviews and references on the subject, we recommend \cite{linder, caldwell09} and~\cite{Schm}.
On a fundamental level,   matter and energy are of quantum nature.
But the usual \qft ~is designed in \minks. Many theoretical and
observational arguments
plead in favour of setting up a rigorous \qft ~in \dS, and of comparing with our familiar minkowskian \qft.
As a matter of fact, the symmetry properties of the de Sitter solutions may allow the
construction of such a theory (see \cite{gazeaurio06,brosmosc} for a~review on the subject). Furthermore, the study of \dS ~\spt ~of\/fers a specif\/ic interest because of   the regularization
opportunity af\/forded by the curvature parameter as a ``natural''
cutof\/f for infrared or other divergences.

On the other hand, some of our most familiar concepts like time, energy,
momentum, etc, disappear. They really require a new conceptual approach in
de Sitterian relativity. However, it should be stressed that the current estimate on the cosmological constant does not allow any palpable experimental ef\/fect on the level of  high energy physics experiments, unless (see \cite{gaznov08}) we deal with theories involving assumptions of inf\/initesimal masses like photon or graviton masses.

As was stressed by Newton and Wigner \cite{newtonwigner}, \emph{the concept of an elementary system $($\dots$)$ is a description of a set of states which forms, in mathematical language, an irreducible representation space for the inhomogeneous Lorentz $({\simeq}$ Poincar\'e$)$ group}. We naturally extend this point of view  by considering  elementary systems in  the de Sitter arena as associated to elements of the unitary dual of $SO_0(1,4)$ or $Sp(2,2)$. The latter was determined a long time ago \cite{thomas,newtondS,dixmier,takahashi} and is compounded of principal, complementary, and discrete series. Note that the de Sitter group has no  unitary irreducible representation (UIR) analogous to the so-called ``massless inf\/inite spin'' UIR of the Poincar\'e group.  As the curvature parameter (or cosmological constant) goes to  zero, some of the de Sitter UIR's have a minskowskian limit that is physically meaningful, whereas the others have not. However, it is perfectly legitimate to study all of them within a~consistent   de Sitter framework, on both mathematical (group representation)  and physical (f\/ield quantization) sides. It should be noticed that some mathematical questions on this unitary dual remain open, like the decomposition of the tensor product of two elements of the discrete series, or should at least be more clarif\/ied, like the explicit realization of representations lying at the lowest limit of the discrete series. Also, the quantization of f\/ields for the latter representations is not known, at the exception of one of them, which is associated with the so-called ``massless minimally coupled f\/ield'' (mmc) in de Sitter\footnote{Note that this current terminology about a certain f\/ield in de Sitter space-time might appear as confusing. In fact  the most general action on a \textit{fixed}, i.e.\ non dynamical, curved space-time that will yield a linear equation of motion for the f\/ield $\phi$ is given by
$
S=\int d^4x \sqrt{-g} \left[ \tfrac{1}{2} g^{\mu \nu} \partial_\mu \phi \partial_\nu \phi-m^2 \phi-\xi R(x) \phi^2\right],
$
where $g_{\mu \nu}$ is the space-time metric, $g=\det g_{\mu \nu}$, and $R(x)$ is the scalar curvature. On an arbitrary curved background, $m$ and $\xi$ are just two real parameters in the theory. In particular the symbol $m$ does not stand for  a physical mass in the minkowskian sense. The equation of motion of this theory is
$
\Box_g \phi+(m^2+\xi R) \phi=0.
$
What is called a  minimally coupled theory  is a theory where $\xi=0$. It is however clear that on a maximally symmetric space-time for which $R(x)=R$ is just a constant the quantity $m^2+\xi R$ alone  really matters.} \cite{gareta,gayou09} and references therein.

The present paper is mainly concerned with this particular family of discrete series of representations. Their carrier spaces present or may display remarkable features: invariant subspace of null-norm states, undecomposable representation features, Gupta--Bleuler triplet, Krein space structure, and underlying cohomology~\cite{pincsimon}. In Section~\ref{background} is given the minimal background to make the reader familiar with de Sitter symmetries and the unitary dual of $Sp(2,2)$. In Section~\ref{contraction1} we give a short  account of the minkowskian content of elements of the unitary dual through group representation contraction procedures. Section~\ref{scalarrep} is devoted to the \emph{scalar}
 representations of the de Sitter group and the associated wave equation. Then, in Section~\ref{genfunct} we construct and control the normalizability of a class of scalar solutions or ``hyperspherical modes'' through de Sitter wave plane solutions \cite{brosgamosc,brosmosc,broseptmosc} viewed as generating functions. The inf\/initesimal and global actions of the de Sitter
group in its scalar unitary representations is described in Section~\ref{dSact}. We then give a	detailed account of the mmc case in Section~\ref{krein}. Finally a list of directions are given in Section~\ref{outlinekrein} in view of future work(s). An appendix is devoted to the root system $B_2$ which corresponds to the de Sitter Lie algebra $so(1,4)$.

\section{de Sitter space-time: geometric and quantum symmetries}\label{background}

We f\/irst recall that the de Sitter space-time is conveniently described as the  one-sheeted hyperboloid embedded in a 4+1-dimensional Minkowski  space, here denoted $\M_5$:
\begin{gather*}
\nonumber M_H \equiv \big\{x \in \M_5 ; \ x^2 := x\cdot x=\eta_{\alpha\beta}  x^\alpha
x^\beta =-H^{2}\big\}, \\
\alpha,\beta  =0,1,2,3,4, \qquad
 \left(\eta_{\alpha\beta}\right)=\mathrm{diag}(1,-1,-1,-1,-1) ,
\end{gather*}
with the so-called ambient coordinates
notations
\[
x := \big(x^0, \vec{x}, x^4\big).
\]

The following  intrinsic coordinates
\begin{gather}\label{coordinates}
	x = \big( x^0=H^{-1}\tan\rho, (H\cos\rho)^{-1} u\big)  ,
	\qquad \rho\in \big]{-}\tfrac{\pi}{2}, \tfrac{\pi}{2}\big[,
	\qquad u \in S^3\,
\end{gather}
are global. They are usually called  ``conformal''.

There exist  ten
Killing vectors  in de Sitterian kinematics. They generate the Lie algebra $so(1,4)$, which gives by exponentiation the de Sitter group $SO_0(1,4)$ or its universal covering $Sp(2,2)$. In unitary irreducible representations of the latter, they are represented as (essentially) self-adjoint operators in  Hilbert
space of (spinor-) tensor valued functions on $M_H$, square integrable with respect to some
invariant inner (Klein--Gordon type) product:
\begin{gather}
\label{gensa}
K_{\alpha \beta} \rightarrow L_{\alpha \beta} = M_{\alpha \beta} + S_{\alpha \beta},
\end{gather}
where
\begin{gather*}
M_{\alpha\beta}=-i(x_{\alpha}\partial_{\beta} - x_{\beta}\partial_{\alpha})
\end{gather*}
 is the ``orbital part'', and
  $S_{\alpha \beta}$ (spinorial part) acts on indices of functions in a certain permutational way.

There are  two Casimir operators, the  eigenvalues  of which  determine  the UIR's:
\begin{gather*}
   \mathcal{C}_2  = - \tfrac{1}{2} L_{\alpha \beta}L^{\alpha \beta}   \qquad (\mbox{quadratic}),\\
 \mathcal{C}_4  = - W_{\alpha} W^{\alpha}, \qquad W_{\alpha} = -
\tfrac{1}{8}\epsilon_{\alpha \beta \gamma \delta \eta} L^{\beta \gamma}L^{\delta \eta} \qquad  (\mbox{quartic}).
\end{gather*}

In a given UIR,  and with the Dixmier notations \cite{dixmier}, these two  Casimir operators are f\/ixed as
\begin{gather}
\label{casim1}
    \mathcal{C}_2  = (-p(p+1) - (q+1)(q-2)) \mathbb{I},   \\
    \mathcal{C}_4  = (-p(p+1)q(q-1)) \mathbb{I} \label{casim2}
\end{gather}
with specif\/ic allowed range of values assumed by parameters $p$ and $q$ for the three series of UIR, namely discrete, complementary, and principal.

\subsubsection*{``Discrete series''  $\boldsymbol{\Pi^{\pm}_{p,q}}$}

Parameter $q$ has a spin meaning. We have to distinguish between
\bei\itemsep=0pt
\item[$(i)$]{\it the scalar case}
$\Pi_{p,0}$, $p=1,2, \dots$. These representations lie at the ``lowest limit'' of the discrete series and  are not square integrable,
\item[$(ii)$] {\it the spinorial case} $\Pi^{\pm}_{p,q}$, $q>0$,
$p= \frac{1}{2}, 1, \frac{3}{2}, 2, \dots$, $q=p, p-1, \dots, 1$ or $\frac{1}{2}$. For $q = \frac{1}{2}$ the representations
$\Pi^{\pm}_{p,\frac{1}{2}}$ are not square-integrable.
\eni

\subsubsection*{``Principal series'' $\boldsymbol{U_{s,\nu}}$}\vspace{-7mm}
\begin{gather*}
  q = \tfrac{1}{2} \pm i \nu.
\end{gather*}
$p=s$ has a spin meaning and  the two Casimir are f\/ixed as
\begin{gather*}
\mathcal{C}_2 = \big(\tfrac{9}{4} + \nu^2 -s(s+1) \big)\mathbb{I} , \qquad
\mathcal{C}_4 =  \big(\tfrac{1}{4}+ \nu^2\big)  s(s+1) \mathbb{I} .
\end{gather*}
We have to distinguish between
\bei\itemsep=0pt
\item[$(i)$] $\nu \in \R$,  $s=1,2, \dots$, for \emph{the integer spin principal series},
\item[$(ii)$] $\nu \neq 0$, $s= \frac{1}{2}, \frac{3}{2}, \frac{5}{2}, \dots$,
 for \emph{the half-integer spin principal series}.
\eni
In both cases, $U_{s,\nu}$ and $U_{s,-\nu}$ are equivalent.
In the case $\nu = 0$, i.e.\ $q = \frac{1}{2}$,   $s= \frac{1}{2}, \frac{3}{2}, \frac{5}{2}, \dots$, the representations are not irreducible. They are direct sums of two UIR's
 belonging to  the discrete series:
 \begin{gather*}
U_{s,0} = \Pi^{+}_{s,\frac{1}{2}} \bigoplus \Pi^{-}_{s,\frac{1}{2}}.
\end{gather*}

\subsubsection*{``Complementary series'' $\boldsymbol{V_{s,  \nu}}$}\vspace{-7mm}
\begin{gather*}
  q = \tfrac{1}{2} \pm \nu .
\end{gather*}
$p=s$ has a spin meaning and  the two Casimir are f\/ixed as
\begin{gather*}
\mathcal{C}_2  =  \big(\tfrac{9}{4} - \nu^2 -s(s+1)\big) \mathbb{I} , \qquad
\mathcal{C}_4  = \big(\tfrac{1}{4} - \nu^2\big) s(s+1)\mathbb{I}.
\end{gather*}
We have to distinguish between
\bei\itemsep=0pt
\item[$(i)$] {\it the scalar case} $V_{0,\nu}$, $\nu \in \R$, $0 < \vert \nu \vert < \frac{3}{2}$,
\item[$(ii)$] {\it the spinorial case} $V_{s,\nu}$,  $0 < \vert \nu \vert < \frac{1}{2}$, $s = 1, 2, 3, \dots$.
\eni
In both cases, $V_{s,\nu}$ and $V_{s,-\nu}$ are equivalent.

\section{Contraction limits or  de Sitterian physics\\ from the point of view of a Minkowskian observer}\label{contraction1}

At this point, it is crucial to understand the physical content of these representations in terms of their null curvature limit, i.e., from the point of view of  local (``tangent'') minkowskian observer, for which the basic physical conservation laws are derived from Einstein--Poincar\'e relativity principles.
We will distinguish between those representations of the de Sitter group which contract to Poincar\'e massive UIR's, those which have a massless content, and those which do not have any f\/lat limit at all.
Firstly let us explain what we mean by null curvature limit on a~geometrical and algebraic level.

\paragraph{On a geometrical level:}
 $\lim\limits_{ H \to 0} M_R = \mathbb{M}_4$, the Minkowski spacetime tangent
to $M_H$ at, say, the de Sitter ``origin'' point $O \deq (0,\vec{0}, H^{-1})$.

\paragraph{On an algebraic  level:}
\begin{itemize}\itemsep=0pt
\item
 $\lim\limits_{ R \to \infty}Sp(2,2) = {\cal P}^{\uparrow}_{+} (1, 3) = \mathbb{M}_4
\SD SL(2,\BC)$, the Poincar\'e group.
\item  The ten de Sitter Killing vectors contract
to their Poincar\'e counterparts $K_{\mu \nu}$, $\Pi_{\mu}$, $\mu =
0, 1, 2, 3$, after rescaling the four $K_{4\mu} \lga \Pi_{\mu} = H
K_{4\mu} $.
\end{itemize}

\subsection{de Sitter UIR contraction: the massive case}\label{dScontm}
For  what we consider as the ``massive'' case, principal series representations  only are
involved  (from which the name  ``de Sitter massive
representations''). Introducing   the Poincar\'e mass $m=\nu/ R$ \cite{micknied,gahure,gaznov08}, we have:
\begin{gather*}
U_{s, \nu} \underset{R \to \infty, \nu \to \infty}{\longrightarrow} {c_>\cal P}^{>}(m,s)
\oplus c_<{\cal P}^{<}(m,s),
\end{gather*}
where one of the ``coef\/f\/icients'' among $c_<$, $c_>$ can be f\/ixed to 1 whilst the other one  vanishes  and where ${\cal P}^{\stackrel{>}{<}}(m,s)$ denotes
the positive (resp.\ negative) energy Wigner UIR's of the Poincar\'e  group with mass $m$ and spin $s$.

\subsection{de Sitter UIR contraction: the massless case}\label{dScontm0}

 Here we must distinguish between
 \begin{itemize}\itemsep=0pt
  \item the scalar massless case, which involves the unique complementary series
UIR $V_{0,1/2}$ to be contractively Poincar\'e signif\/icant,

  \item and the helicity $s\neq 0$ case where are involved all representations
$\Pi^{\pm}_{s,s}$, $s>0$ lying at the lower limit of the discrete
series.
\end{itemize}

The arrows $\hookrightarrow $ below designate unique
extension. Symbols ${\cal P}^{\overset{>}{ <}}(0,s)$ denote the Poincar\'e massless representations
with helicity $s$ and with positive (resp. negative) energy. Conformal
invariance involves  the discrete series
representations
(and their lower limits) of the (universal covering of the)
conformal group or its double covering $SO_0(2,4)$ or its fourth
covering $SU(2,2)$. These UIR's are denoted  by
${\cal C}^{\stackrel{>}{<}}(E_0,j_1, j_2)$, where $(j_1,j_2) \in
\N/2 \times \N/2$ labels the UIR's of $SU(2) \times SU(2)$ and
$E_0$ stems for the positive (resp. negative) conformal energy.
\bei\itemsep=0pt
\item Scalar massless case:
\begin{gather*}
\nonumber \left. \begin{array}{ccccccc}
& & {\cal C}^{>}(1,0,0)
& &{\cal C}^{>}(1,0,0) &\hookleftarrow &{\cal P}^{>}(0,0)\\
V_{0,1/2} &\hookrightarrow & \oplus
&\stackrel{R \to \infty}{\longrightarrow} & \oplus & &\oplus \\
& & {\cal C}^{<}(-1,0,0)&
& {\cal C}^{<}(-1,0,0) &\hookleftarrow &{\cal P}^{<}(0,0).\\
\end{array} \right.
\end{gather*}
\item Spinorial massless case:

\begin{gather*}
\nonumber \left. \begin{array}{ccccccc}
& & {\cal C}^{>}(s+1,s,0)
& &{\cal C}^{>}(s+1,s,0) &\hookleftarrow &{\cal P}^{>}(0,s)\\
\Pi^+_{s,s} &\hookrightarrow & \oplus
&\stackrel{R \to \infty}{\longrightarrow} & \oplus & &\oplus \\
& & {\cal C}^{<}(-s-1,s,0)&
& {\cal C}^{<}(-s-1,s,0) &\hookleftarrow &{\cal P}^{<}(0,s),\\
\end{array} \right. \end{gather*}
\begin{gather*}
\nonumber \left. \begin{array}{ccccccc}
& & {\cal C}(s+1,0,s)
& &{\cal C}^{>}(s+1,0,s) &\hookleftarrow &{\cal P}^{>}(0,-s)\\
\Pi^-_{s,s} &\hookrightarrow & \oplus
&\stackrel{R \to \infty}{\longrightarrow} & \oplus & &\oplus \\
& & {\cal C}^{<}(-s-1,0,s)&
& {\cal C}^{<}(-s-1,0,s) &\hookleftarrow &{\cal P}^{<}(0,-s).\\
\end{array} \right. \end{gather*}
\eni

\section{Scalar representations}\label{scalarrep}

In the present study, we are  concerned  with scalar f\/ields only, for which the value of the quartic Casimir vanishes. Two cases are possible: $p=0$ for the principal and complementary series, and $q=0$ for the discrete series. In both cases, the f\/ields carrying the representations are solutions of the scalar quadratic ``wave equations'' issued from  equation  (\ref{casim2}):
\begin{gather}
\label{scalwavdis}
\mathcal{C}_2 \psi (x) \equiv Q_0 \psi (x) = -(p-1)(p+2)\psi (x),
\end{gather}
for the scalar discrete series, and
\begin{gather}
\label{scalwavpc}
\mathcal{C}_2 \psi (x) \equiv Q_0 \psi (x) = -(q+1)(q-2)\psi (x) ,
\end{gather}
for the scalar principal and complementary series.
Let us  def\/ine the symmetric, ``transverse projector''
\begin{gather*}
\theta_{\alpha \beta}= \eta_{\alpha \beta}+H^2x_\alpha x_\beta
\end{gather*}
which satisf\/ies $\theta_{\alpha\beta}
x^{\alpha}=\theta_{\alpha\beta}  x^{\beta}=0$. It is the
transverse form of the de Sitter metric in ambient space notations and it is used in the construction of  transverse entities like  the
transverse derivative
\begin{gather*}
\bar
\partial_\alpha=\theta_{\alpha \beta}\partial^\beta=
\partial_\alpha+H^2x_\alpha x . \partial  .
\end{gather*}

\begin{figure}[th]
\centerline{\includegraphics{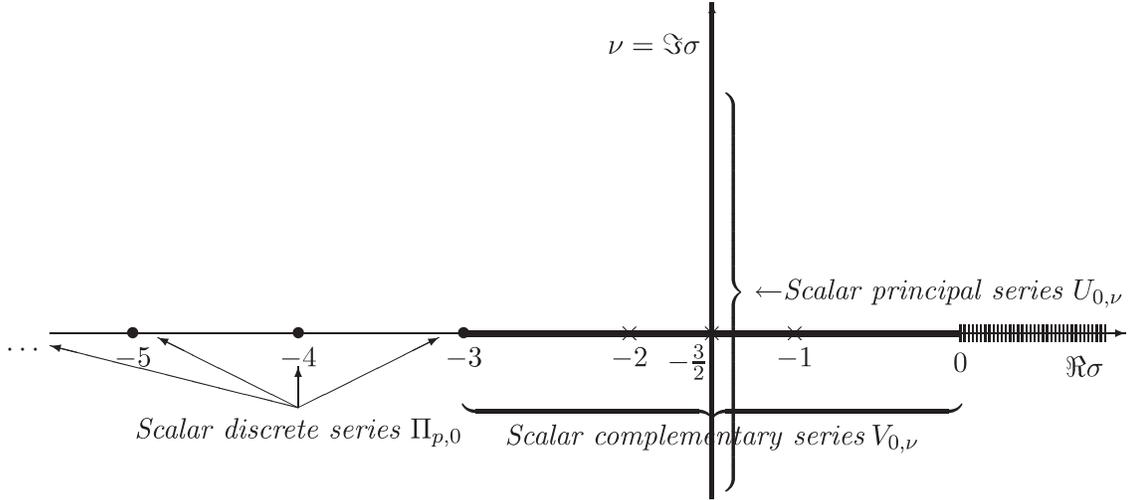}}

%
%

\caption{Indexing the set of scalar UIR with  complex parameter $\sigma$. It should be remembered that representations $U_{0,\nu}$ and $U_{0,-\nu}$ with $\sigma = -3/2 -i\nu$  in the principal series, and  $V_{0,\nu }$ and $V_{0,-\nu}$ with $\sigma = -3/2 -\nu$ in the complementary series (i.e. under symmetry $\sigma \mapsto -3 - \sigma$)  are equivalent. The ``massless minimally coupled'' representation corresponds to $\sigma = -3$. The ``conformally invariant massless'' scalar representation corresponds to $\sigma = -2$, or equivalently to $\sigma = -1$. The case $\sigma = 0$ can be viewed  as corresponding to the trivial representation.}
\label{scalrep}
\end{figure}

With these notations, the scalar Casimir operator reads as $Q_0 = -H^{-2}\bar\partial^2 $ and equations~(\ref{scalwavdis}) and (\ref{scalwavpc}) become
\begin{gather}
\label{wavscal}
(Q_0+ \sigma(\sigma + 3))\psi(x)= -H^{-2} \bar
\partial^2 \psi(x) + \sigma(\sigma + 3)\psi(x) = 0,
\end{gather}
where we have introduced the unifying complex parameter $\sigma$. As is shown in Fig.~\ref{scalrep},  $\sigma = p-1$ or $=-p-2$ for the scalar discrete series,  $\sigma = -q-1= -3/2 -i\nu$ for  the scalar principal series, and  $\sigma = -q-1= -3/2 -\nu$ for the scalar complementary series. Actually, we will examine this equation for any complex value of the parameter $\sigma$, proceeding with appropriate  restrictions when is necessary. Just note that the scalar discrete series starts with the so-called massless minimally coupled'' (mmc) case, exactly there where the complementary series ends on its left.

\section{de Sitter wave planes as generating functions}\label{genfunct}

 There exists a continuous family of simple solutions (``de Sitter plane waves'')  to equation (\ref{wavscal}). These solutions,  as   indexed by vectors $\xi$ lying in  the null-cone in $\M_5$:
 \begin{gather*}
\xi = (\xi^0, \pmb{\xi}) \in\mathcal{C} \deq\left\lbrace  \xi \in \M_5  : \xi^2 = 0\right\rbrace   ,
\end{gather*}
read
\begin{gather*}
\psi(x)=(Hx\cdot \xi)^{\sigma} .
\end{gather*}
Putting $\pmb{\xi} = \Vert \pmb{\xi} \Vert v \in \mathbb{R}^4$, $v \in S^3$, and $\vert \xi^0 \vert = \Vert \pmb{\xi} \Vert$, we rewrite  the dot product $Hx\cdot \xi$ as follows{\samepage
\begin{gather*}
Hx\cdot \xi  = (\tan{\rho})\,\xi^0 - \frac{1}{\cos{\rho}} u\cdot\pmb{\xi}
 = \frac{\xi^0 e^{i\rho}}{2i\,\cos{\rho}}\big(1 + z^2 - 2zt\big)\\
\nonumber  \phantom{Hx\cdot \xi  =}{} \mbox{with}\ z = i e^{- i\rho}\mathrm{sgn}\,{\xi^0}, \qquad t = u\cdot v =: \cos{\varpi} .
\end{gather*}}

By using  the generating function for  Gegenbauer polynomials,
\begin{gather}
\label{expgeg}
\big(1 + z^2 - 2 z t\big)^{-\lambda} = \sum_{n=0}^{\infty} z^n \, C_n^{\lambda}(t), \qquad \vert z \vert <1 ,
\end{gather}
we get the expansion
\begin{gather}
\label{expgegen}
(Hx\cdot\xi)^{\sigma} =  \left(\frac{\xi^0 e^{i\rho}}{2i\,\cos{\rho}}\right)^{\sigma} \big(1 + z^2 - 2zt\big)^{\sigma}= \left(\frac{\xi^0 e^{i\rho}}{2i\,\cos{\rho}}\right)^{\sigma}  \sum_{n=0}^{\infty} z^n  C_n^{-\sigma}(t) , \qquad \Re{\sigma} < \frac{1}{2}  .
\end{gather}
This expansion is actually not valid  in the sense of  functions since $\vert z \vert = 1$. However, giving a~negative imaginary part to the angle $\rho$ ensures the convergence. This amounts to extend ambient coordinates to the forward tube \cite{brosmosc}:
\begin{gather*}
	\mathcal{T}^{+} = \big\{ \M_5 - i \overline{V}_5^{+} \cap M_H^{\scriptscriptstyle\mathbb{C}} \big\},
	\qquad \overline{V}_5^{+} = \big\{ x \in \M_5 : x^2 \geq 0, x^0 > 0 \big\}.	
\end{gather*}

We now make use of two expansion formulas involving Gegenbauer polynomials \cite{hua} and $S^3$ normalized hyperspherical harmonics:
\begin{gather}
  C_n^{\lambda}(t)  =
  \frac{1}{\Gamma(\lambda)\Gamma(\lambda-1)} \sum_{k = 0}^{\lfloor \frac{n}{2}\rfloor} c_k   C_{n-2k}^{1} (t) , \nonumber\\
  c_k = \frac{(n-2k + 1) \Gamma(k+\lambda - 1) \Gamma(\lambda + n - k)}{k! \Gamma(n-k + 2)}   ,\label{gegen}\\
 \label{geghyp}  C_L^1(v\cdot v') =    \frac{2\pi^2}{L+1}\sum_{lm} \mathrm{Y}_{Llm}(v) \mathrm{Y}^{\ast}_{Llm}(v') , \qquad v, v' \in S^3 .
\end{gather}

We recall here the expression of the hyperspherical harmonics:
\begin{gather}
\label{hypersph}
\mathrm{Y}_{Llm}(u)
=\left(\frac{(L+1)(2l+1)(L-l)!}{2\pi^2(L+l+1)!}\right)^{\frac{1}{2}}
2^ll!\left(\sin\alpha\right)^lC_{L-l}^{l+1}\left(\cos\alpha\right)
Y_{lm}(\theta,\phi) ,
\end{gather}
for $(L,l,m)\in\N\times\N\times\Z$ with $0\leq l\leq L$ and $-l\leq
m\leq l$. In this equation the $Y_{lm}$'s are ordinary spherical harmonics:
\[
Y_{lm}(\theta,\phi)=
(-1)^m\left(\frac{(l-m)!}{(l+m)!}\right)^{\frac{1}{2}}
P_l^m(\cos\theta)e^{im\phi},
\]
where the $P_l^m$'s are the associated Legendre functions.
With this choice of constant factors, the $\mathrm{Y}_{Llm}$'s obey the
orthogonality (and normalization) conditions:
\[
\int_{S^3}\mathrm{Y}_{Llm}^{*}(u)\mathrm{Y}_{L'l'm'}(u)\,du
=\delta_{LL'}\delta_{ll'}\delta_{mm'}.
\]

Combining \eqref{expgeg},  \eqref{gegen} and \eqref{geghyp} we get the expansion formula:
\begin{gather}
\label{geghyp1}
\big(1 + z^2 - 2 z v\cdot v'\big)^{-\lambda} = 2\pi^2 \sum_{Llm} z^{L}p^{\lambda}_{L}\big(z^2\big)\mathrm{Y}_{Llm}(v) \mathrm{Y}^{\ast}_{Llm}(v'),
\end{gather}
where
\begin{gather*}
	p^{\lambda}_{L}\big(z^2\big)   =
	\frac{1}{(L+1)!}\frac{\Gamma(\lambda+L)}{\Gamma(\lambda)}
	{}_2F_1\big(L+\lambda, \lambda - 1 ; L + 2 ; z^2\big)
\end{gather*}
and the integral representation,
\begin{gather*}
z^{L}p^{\lambda}_{L}\big(z^2\big)  \mathrm{Y}_{Llm}(v) = \frac{1}{2\pi^2} \int_{S^3} \big(1 + z^2 - 2 z v\cdot v'\big)^{-\lambda} \mathrm{Y}_{Llm}(v') d\mu(v') .
\end{gather*}

Let us  apply the above material to the de Sitter plane waves $(Hx\cdot \xi )^{\sigma}$. In view of equations~(\ref{expgegen}) and (\ref{geghyp1}) with $\lambda = -\sigma$, we introduce the  set of functions on the de Sitter hyperboloid:
\begin{gather}
\nonumber \Phi_{Llm}^{\sigma} (x)= \frac{i^{L-\sigma} e^{-i(L-\sigma)\rho}}{(2\cos{\rho})^{\sigma}} p^{-\sigma}_{L}\big({-}e^{- 2i\rho}\big) \mathrm{Y}_{Llm}(u)\\
\phantom{\Phi_{Llm}^{\sigma} (x)}{}
= \frac{i^{L-\sigma} e^{-i(L-\sigma)\rho}}{(2\cos{\rho})^{\sigma}}\frac{\Gamma(L-\sigma)}{(L+1)!\Gamma(-\sigma)}\, {}_2F_1\big(L-\sigma,-\sigma - 1; L+2;-e^{- 2i\rho}\big) \mathrm{Y}_{Llm}(u) .\!\!\label{setbasisdS}
\end{gather}
By using the well-known relation between hypergeometric functions \cite{magnus},
\begin{gather*}
{}_2F_1(a,b;c;z) = (1-z)^{c-a-b}\, {}_2F_1(c-a,c-b;c;z) ,
\end{gather*}
we get the alternative form of (\ref{setbasisdS})
\begin{gather}
 \Phi_{Llm}^{\sigma} (x) = i^{L-\sigma}\, e^{-i(L+\sigma +3)\rho}(2 \cos{\rho})^{\sigma + 3} \frac{\Gamma(L-\sigma)}{(L+1)!\Gamma(-\sigma)} \nonumber\\
\phantom{\Phi_{Llm}^{\sigma} (x) =}{}  \times {}_2F_1\big(\sigma + 2, L+\sigma +3; L+2;-e^{- 2i\rho}\big)  \mathrm{Y}_{Llm}(u) .\label{setbasisdS1}
\end{gather}
We then have the expansion of the de Sitter plane waves:
\begin{gather*}
(Hx\cdot \xi )^{\sigma} =  2\pi^2   \sum_{Llm}  \Phi_{Llm}^{\sigma} (x)   \big(\xi^0\big)^{\sigma}  \big(\mathrm{sgn}\,{\xi^0}\big)^L  \mathrm{Y}^{\ast}_{Llm}(v)  .
\end{gather*}
From the linear independence of the hyperspherical harmonics, it is clear that the functions $\Phi_{Llm}^{\sigma} (x)$ are solutions to the scalar wave equation (\ref{wavscal}) once one has proceeded with the appropriate separation of variables, a question that we examine in the next section.
From the orthonormality of the set of hyperspherical harmonics we have the integral representation (``Fourier transform'' on~$S^3$),
\begin{gather*}
\Phi_{Llm}^{\sigma} (x) =  \frac{\big(\mathrm{sgn}\,{\xi^0}\big)^L}{2\pi^2  (\xi^0)^{\sigma} }  \int_{S^3} d\mu(v)  (Hx\cdot \xi )^{\sigma}    \mathrm{Y}_{Llm}(v)  .
\end{gather*}
We notice that the functions $\Phi_{Llm}^{\sigma} (x)$ are well def\/ined for all $\sigma$ such that $\Re\sigma < 0$, so for all scalar de Sitter UIR,  and are inf\/initely dif\/ferentiable in  the conformal coordinates $(\rho,u)$ in their  respective ranges. At the inf\/inite de Sitter ``past'' and ``future'', i.e.\ at the limit $\rho = \pm \pi/2$, their behavior is ruled by the factor $(\cos{\rho})^{\sigma + 3}$:
\begin{gather*}
\Phi_{Llm}^{\sigma} (x) \simeq_{\rho \to \pm \pi/2}   i^{L-\sigma} e^{-i(L+\sigma +3)\rho}(2\,\cos{\rho})^{\sigma + 3}\frac{\Gamma(-2\sigma -1)}{(L+1)!\Gamma(-\sigma)\Gamma(-\sigma -1)} \mathrm{Y}_{Llm}(u) ,
\end{gather*}
where we have used the formula \cite{magnus}
\[
{}_2F_1(a,b;c;1)= \frac{\Gamma(c)\Gamma(c-a-b)}{\Gamma(c-a)\Gamma(c-b)}
\]
 valid for $\Re(c-a-b) >0$ and $c \neq -1,-2,\dots$. The singularity that appears for $\Re\sigma < -3$, which is the case for the scalar discrete series with $p \geq 2$, is due to the choice of conformal coordinates in expressing the dot product $\xi\cdot x$.

We now ask the question about the  nature of the above functions as basis elements of some specif\/ic  vector space of solutions to equation~(\ref{wavscal}). For that we f\/irst introduce the so-called  Klein--Gordon inner product in the space of solutions to~(\ref{wavscal}),
def\/ined for solutions  $\Phi_1$, $\Phi_2$ by
\begin{gather}
\label{kgie}
 \langle\Phi_1,\Phi_2\rangle=i\int_\Sigma \Phi_1^{*}(x)\big(\overrightarrow{
   \partial}_{\mu}-\overleftarrow{\partial}_{\mu}\big)\Phi_2(x)\,d\sigma^\mu\equiv
i\int_\Sigma \Phi_1^*\stackrel{\leftrightarrow}{
   \partial}_{\mu}\Phi_2 \,d\sigma^\mu,
\end{gather}
where $\Sigma$ is a Cauchy surface, i.e.\ a space-like surface such that the Cauchy
data on $\Sigma$ def\/ine uniquely a solution of~(\ref{wavscal}), and $d \sigma^\mu$
is the area element vector on $\Sigma$. This product is de Sitter invariant and
independent of the choice of $\Sigma$.
In accordance with our choice of global coordinate
system, the Klein--Gordon  product~(\ref{kgie}) reads as
 \begin{gather*}
 \langle\Phi_1,\Phi_2\rangle=\frac{i}{H^2}\int_{\rho=0}
 \Phi_1^{*}(\rho,u)\stackrel{\leftrightarrow}{
\partial}_{\rho}\Phi_2(\rho,u)\,
du  ,\end{gather*}
where $du=\sin^2\alpha \sin\theta\,d\alpha\,d\theta\,d\phi$ is
the invariant measure on $S^3$.
Due to the orthogonality of the hyperspherical harmonics, the set of functions $\Phi_{Llm}^{\sigma} (x) $ is orthogonal:
\begin{gather*}
\langle\Phi_{L_1l_1m_1}^{\sigma},\Phi_{L_2l_2m_2}^{\sigma}\rangle = \delta_{L_1L_2}  \delta_{l_1l_2}  \delta_{m_1m_2}  \Vert \Phi_{L_1l_1m_1}^{\sigma}\Vert^2  ,
\end{gather*}
   in case of normalizable states. Let us calculate this norm:
\begin{gather}
\Vert \Phi_{Llm}^{\sigma}\Vert^2 = \frac{ \pi\, 2^{2-2L}}{H^2}e^{-\pi \Im\sigma}\left\vert \frac{(\Gamma(L-\sigma)}{(\Gamma(-\sigma)}\right\vert^2 \nonumber\\
\phantom{\Vert \Phi_{Llm}^{\sigma}\Vert^2 =}{} \times  \Re\left\lbrack \Gamma^{\ast}\left(\frac{L-\sigma+1}{2}\right)
\Gamma\left(\frac{L-\sigma}{2}\right)\Gamma^{\ast}\left(\frac{L+\sigma+4}{2}\right)
\Gamma\left(\frac{L+\sigma+3}{2}\right)\right\rbrack^{-1}  .\label{normPhicompl}
\end{gather}
For real values of $\sigma$, in particular for the complementary series and, with restrictions for the discrete series (see below), the norm simplif\/ies to:
\begin{gather}
\label{normPhireal}
\Vert \Phi_{Llm}^{\sigma}\Vert^2=\frac{ 2^3}{H^2}\frac{\Gamma(L-\sigma)}{(\Gamma(-\sigma))^2\Gamma(L+\sigma + 3)} .
\end{gather}
We see that for the scalar principal and complementary series all these functions are normalizable and are suitable candidates for scalar f\/ields in de Sitter space-time carrying their respective UIR.

For the discrete series, $\sigma = -p-2$, $p=1,2,\dots,$ the hypergeometric functions reduce to polynomials of degree~$p$, ${}_2F_1(-p, L- p +1; L+2;-e^{- 2i\rho})$, and the norm vanishes for states with $L=0, 1, \dots, p-1$. We will interpret this invariant $N$-dimensional null-norm subspace, $N = p(p+1)(2p+1)/6$, as a space of ``gauge'' states, carrying the irreducible (non-unitary!) de Sitter f\/inite-dimensional representation  $(n_1=0,n_2 = p-1)$ (with the notations of Appendix~\ref{B2}), which is ``Weyl equivalent'' to the UIR $\Pi_{p,0}$, i.e.\ shares with it the same eigenvalue of the Casimir operator $\mathcal{C}_2$.

For  the regular case $L \geq p$,  one can  re-express the $\rho$-dependent part of the functions $\Phi_{Llm}^{\sigma}\equiv \Phi_{p;Llm}$ in terms of Gegenbauer polynomials:
\begin{gather*}
\Phi_{p;Llm} (x)= -i^{L+p} 2^{1-p}  \frac{(L+p+1)!  (L-p)!}{(L!)^2  (L+1)(p+1)}\nonumber\\
\phantom{\Phi_{p;Llm} (x)=}{}\times e^{-i(L+1)\rho}(\cos{\rho})^{1-p} C^{L-p+1}_p(\sin{\rho})  \mathrm{Y}_{Llm}(u)  .
\end{gather*}

In the allowed ranges of parameters, the normalized functions \cite{ChernTag} are def\/ined as:
\begin{gather}
	\Psi_{Llm}^\sigma(x) =
	\mathcal{N}_L^\sigma i^{L-\sigma}
	e^{-i(L+\sigma +3)\rho}(\cos\rho)^{\sigma + 3}\,
	{}_2F_1(\sigma +1, L+\sigma + 3;  L+ 2; -e^{-2i\rho}) \mathrm{Y}_{Llm}(u),\nonumber\\
	\mathcal{N}_L^\sigma  = \frac{H}{\sqrt{\pi}}  2^{L + \sigma +2}  e^{\frac{\pi}{2}\Im\sigma}  \frac{\Gamma(L-\sigma)}{\vert \Gamma(L-\sigma)\vert}  \frac{\vert\Gamma(-\sigma)\vert}{\Gamma(-\sigma)} \frac{1}{(L+1)!} \nonumber\\
\phantom{\mathcal{N}_L^\sigma  =}{} \times \left\lbrack \Re\left( \Gamma^{\ast}\left(\frac{L-\sigma+1}{2}\right)
\Gamma\left(\frac{L-\sigma}{2}\right)
\Gamma^{\ast}\left(\frac{L+\sigma+4}{2}\right)\Gamma\left(\frac{L+\sigma+3}{2}\right)\right)\right\rbrack^{1/2} .
\end{gather}
 In the complementary and discrete series, $\sigma$ is real and, due to the duplication formula for the gamma function,  the expression between brackets reduces to $\pi 2^{2-2L-3}\Gamma(L-\sigma) \Gamma(L +\sigma +3)$. Then the normalization factor simplif\/ies to:
 \begin{gather*}
\mathcal{N}_L^\sigma = H  2^{\sigma + 1/2}  \frac{\sqrt{\Gamma(L-\sigma) \Gamma(L +\sigma +3)}}{(L+1)!}  .
\end{gather*}
Finally, in the scalar discrete series, with $\sigma = -p-2$, one gets for $L\geq p$ the orthonormal system:
\begin{gather*}
 	\Psi_{Llm}^\sigma(x) \equiv \Psi_{p;Llm}(x)  =
	-\mathcal{N}_{p;L}
	i^{L+p} e^{-i(L+1 -p) \rho}(\cos\rho)^{1-p} \nonumber\\
	\phantom{\Psi_{Llm}^\sigma(x) \equiv \Psi_{p;Llm}(x)  =}{}
 \times{}_2F_1\big(-p, L+1-p; L+2; -e^{-2i\rho}\big) \mathrm{Y}_{Llm}(u)  ,\\ 
\mathcal{N}_{p;L}  =
	2^{-p-1/2}
	\frac{\sqrt{\Gamma(L-p+1)\Gamma(L+p+2)}}{(L+1)!}  , \qquad L \geq p  .
\end{gather*}
As noticed above,  those functions become singular at the limits $\rho = \pm \pi/2$ at the exception of the lowest case (``minimally coupled massless f\/ield'') $p=1$.
Going back to the de Sitter plane waves as generating functions, one gets the expansion in terms of orthonormal sets for the scalar principal or complementary series:
\begin{gather*}
	(Hx\cdot\xi)^\sigma = 2\pi^2\sum_{Llm}\Vert \Phi^\sigma_{Llm}\Vert \Psi^\sigma_{Llm}(x)
	\big(\xi^0\big)^\sigma \big(\mathrm{sgn}\, \xi^0\big)^L Y^\ast_{Llm}(v) ,
\end{gather*}
where $\Vert \Phi^\sigma_{Llm}\Vert $ is given by (\ref{normPhicompl}) (principal) and by  (\ref{normPhireal}) (complementary).
For the scalar discrete series, we have to split the sum into two parts:
\begin{gather*}
\nonumber	(Hx\cdot\xi)^{-p-2}  = 2\pi^2   \sum_{L=0}^{p-1}\sum_{lm}  \Phi_{p;Llm} (x)   \big(\xi^0\big)^{-p-2} \big(\mathrm{sgn}\,{\xi^0}\big)^L \mathrm{Y}^{\ast}_{Llm}(v)  \\
\phantom{(Hx\cdot\xi)^{-p-2}  =}{}
+  2\pi^2\sum_{L=p}^{\infty}\sum_{lm}
	\Vert \Phi_{p;Llm}\Vert \Psi_{p;Llm}(x)
	\big(\xi^0\big)^{-p-2} \big(\mathrm{sgn} \xi^0\big)^L Y^\ast_{Llm}(v)  .
\end{gather*}
These formulae make explicit the ``spherical'' modes in de Sitter space-time in terms of de Sitter plane waves.

\section{Wave equation for scalar de Sitter representations}

Let us check how we recover the functions $\Phi^{\sigma}_{Llm}$ or $\Psi^{\sigma}_{Llm}$ by directly solving the wave equation.
The scalar Casimir operator $Q_0$ introduced in (\ref{scalwavdis}) and (\ref{scalwavpc}) is just proportional to the Laplace--Beltrami operator on de Sitter space: $Q_0 = -H^{-2}\square$.
 In terms of the conformal coordinates~(\ref{coordinates}) the latter is given by
 \begin{gather*}
\square=\frac{1}{\sqrt g}\partial _\nu\sqrt gg^{\nu\mu}\partial _\mu=
   H^2  \cos^4 \rho \frac{\partial }{\partial \rho}\left(\cos^{-2} \rho
    \frac{\partial }{\partial \rho}\right)-H^2\cos^2 \rho  \Delta_{3}  ,
\end{gather*}
where
\begin{gather*}
\Delta_{3}=\ddpx{\alpha}+2\cot\alpha\dpx{\alpha}+\frac{1}{\sin^2\alpha}
\ddpx{\theta}+\cot\theta\frac{1}{\sin^2\alpha}\dpx{\theta}
+\frac{1}{\sin^2\alpha\sin^2\theta}\ddpx{\phi}
\end{gather*}
 is the Laplace operator on the
hypersphere~$S^3$.
Equation (\ref{scalwavpc}) can be solved by separation of variable
\cite{ChernTag,kirga}. We put
\[
\psi(x)=\chi(\rho)D(u),
\]
where $u\in S^3$,
and obtain
      \begin{gather} [\Delta_{3}+C]D(u) = 0,\label{eqang}  \\
 \left(\cos^4 \rho \frac{d}{d\rho}\cos^{-2} \rho \frac{d}{d\rho}
          +C\cos^2 \rho
-\sigma(\sigma + 3)\right)\chi(\rho) = 0.\label{eqrad}
\end{gather}

We begin with the angular part problem~(\ref{eqang}). For
$C=L(L+2)$, $L\in\N$ we f\/ind the hyperspherical harmonics
$D=\mathrm{Y}_{Llm}$  which are def\/ined in~(\ref{hypersph}).

For the $\rho$ dependent part, and for $\sigma$ we obtain the solutions in terms of Legendre functions on the cut
 \begin{gather*}
\chi_{\sigma; L}(\rho)=A_L(\cos
      \rho)^{\frac{3}{2}}\left\lbrack P^{\lambda}_{L+\frac{1}{2}}(\sin\rho)-
\frac{2i}{\pi}
    Q^{\lambda}_{L+\frac{1}{2}}(\sin\rho)\right\rbrack.
\end{gather*}
Here $\lambda = \pm (\sigma + 3/2)$ and $A_L$ is given by
\[
 A_L=H\frac{\sqrt\pi}{2}\left(\frac{\Gamma\big(L-\lambda+\frac{3}{2}\big)}
    {\Gamma\big(L+\lambda+\frac{3}{2}\big)}\right)^{\frac{1}{2}}.
 \]
We then obtain the  set of solutions
\begin{gather*}
\Psi_{Llm}^\sigma(x)=\chi_{\lambda L}(\rho)\mathrm{Y}_{Llm}(u),\qquad  x=(\rho,u)\in M_H,
\end{gather*}
for the f\/ield equation $(\square-\sigma(\sigma +3))\psi=0$
Note that this family of solutions is orthonormal for the scalar complementary series and for the discrete series in the allowed range. In the discrete series and for $L \leq p-1$, the null-norm states $\Phi_{p;Llm}$'s are orthogonal to all other elements $\Phi_{p;L'l'm'}$, whatever their normalizability. All elements  satisfy also another orthogonality relation:
\begin{gather*}
\langle\Phi_{L'l'm'}^\sigma,\left(\Phi_{Llm}^\sigma\right)^{*}\rangle=0  .
\end{gather*}
The link with the hypergeometric functions appearing in (\ref{setbasisdS}) can be found directly from the explicit expansions of the Legendre functions in their arguments. It can be also found from the dif\/ferential equation~(\ref{eqang}) through the change of variables $\rho \mapsto z = -e^{-2i\rho}$, $\chi(\rho) = h(z)$:
\begin{gather}
\label{eqinz}
\left(z^2(1-z)^2\frac{d^2}{dz^2} + 2 z(1-z)\frac{d}{dz} - \frac{1}{4}L(L+2) (1-z)^2 - z\sigma(\sigma +3)\right) h(z) = 0  ,
\end{gather}

\paragraph{Frobenius solutions in the neighborhood of $\boldsymbol{z=0}$.}
The Frobenius indicial equation for solutions of the type $z^c\sum\limits_{n\geq 0} a_nz^n$ has two solutions: $c= c_1 = L/2$, which corresponds to what we got in (\ref{setbasisdS}) or (\ref{setbasisdS1}), and $c=c_2=-(L+1)/2$, i.e. a f\/irst solution is given by
\begin{gather}
\label{regsol}
h(z) = h_1(z) = z^{\frac{L}{2}}  (1-z)^{\sigma +3}\, {}_2F_1(\sigma +2,L+\sigma +3;L+2;z) .
\end{gather}
Since $c_1-c_2 = L+1 \in \N$, we have to deal with to   degenerate case, which means that a linearly independent solution has the form
\begin{gather}
\label{singsol111}
h(z) = h_2(z) = (\log{z})   h_1(z) + \sum_{n=-L-1}^{+\infty} b_n  z^{n + L/2}  ,
\end{gather}
where coef\/f\/icients $b_n$ are recurrently determined from~(\ref{eqinz}).
This second set of solutions takes all its importance in the discrete case when we have to deal with the f\/inite dimensional space of null-norm solutions, as is shown  in Section~\ref{krein} for the simplest case $p=1$.
 The respective Klein--Gordon norms of these solutions are given by
\begin{gather}
\label{kgnormreg2}
\|h_1\|^2 =\frac{2^3}{H^2}\frac{(L+1)!^2}{\Gamma(L-\sigma)\Gamma(L+\sigma+3)}  ,
\end{gather}
which corresponds to (\ref{normPhicompl}), and (for $b_n$ real)
\begin{gather}
\label{kgnormsing2}
 \|h_2\|^2 =\pi^2 \|h_1\|^2 + \frac{u}{H^2}\frac{\sqrt{\pi } 2^{2-L} (L+1)!}{\Gamma \left(\frac{L-\sigma+1}{2} \right) \Gamma \left(\frac{L+\sigma +4}{2}\right)}+(-1)^L \frac{4uv}{H^2} ,
\end{gather}
where we have introduced the following quantities
\[
u=\sum_{n=-L-1}^\infty (-1)^n b_n, \qquad v=\sum_{n=-L-1}^\infty (-1)^n (n+L/2)b_n .
\]
We conjecture  that in the cases $\sigma = -p-2$, $0\leq L \leq p-1$, the norms (\ref{kgnormsing2}) vanish like the norms (\ref{kgnormreg2}).

\paragraph{Frobenius solutions in the neighborhood of $\boldsymbol{z=1}$ for the discrete series.}
The  indicial equation for solutions of the type $(1-z)^d\sum\limits_{n\geq 0} c_n(1-z)^n$ to equation~(\ref{eqinz}) has two solutions: $d= d_1 = -\sigma = p+2$, and $d=d_2=\sigma +2 = 1-p$. The latter corresponds to  what we got in~(\ref{setbasisdS}) or (\ref{setbasisdS1}). The former gives the following regular solution in the neighborhood of $z=1$:
\begin{gather*}
w(z) = w_1(z) = z^{\frac{L}{2}}  (1-z)^{p+2}\, {}_2F_1(p+1,L+ p +2;L+2;1-z) ,
\end{gather*}
or, in term of hypergeometric polynomial,
\begin{gather*}
w(z) = w_1(z) = z^{\frac{L}{2}-2p -1}  (1-z)^{p+2}\, {}_2F_1(-p,L- p +1;L+2;1-z)  .
\end{gather*}
Since $d_1-d_2 = 2p+1 \in \N$, we are again in presence of a    degenerate case. A linearly independent solution is given by
\begin{gather*}
w(z) = w_2(z) = \log{(1-z)}   w_1(z) + \sum_{n=-2p-1}^{+\infty} e_n  (1-z)^{n + p+2}  ,
\end{gather*}
where coef\/f\/icients $e_n$ are recurrently determined from  (\ref{eqinz}) after change of variables $z \mapsto 1-z$.

\section{de Sitter group actions}\label{dSact}

\subsection{Inf\/initesimal actions}\label{infdSact}
Let us express the inf\/initesimal generators  (\ref{gensa})  in terms of conformal  coordinates.

 The six generators of the compact $SO(4)$
subgroup, contracting to the Euclidean subalgebra when $H\to 0$,   read as follows
\begin{gather*}
M_{12} =-i\dpx{\phi},\\
M_{32} =-i\left(\sin\phi\dpx{\theta}+\cot\theta\cos\phi\dpx{\phi}\right),\\
M_{31} =-i\left(\cos\phi\dpx{\theta}+\cot\theta\sin\phi\dpx{\phi}\right),\\
M_{41} =-i\left(\sin\theta\cos\phi\dpx{\alpha}
+\cot\alpha\cos\theta\cos\phi\dpx{\theta}
-\cot\alpha\frac{\sin\phi}{\sin\theta}\dpx{\phi}\right),\\
M_{42} =-i\left(\sin\theta\sin\phi\dpx{\alpha}
+\cot\alpha\cos\theta\sin\phi\dpx{\theta}
+\cot\alpha\frac{\cos\phi}{\sin\theta}\dpx{\phi}\right),\\
M_{43} =-i\left(\cos\theta\dpx{\alpha}
-\cot\alpha\sin\theta\dpx{\theta}\right).
\end{gather*}
The four generators contracting  to  time translations and Lorentz boosts
 when $H\to 0$ read as follows
\begin{gather*}
M_{01} =-i\left(\cos\rho\sin\alpha\sin\theta\cos\phi\dpx{\rho}
 \right.\\
  \left.\phantom{M_{01} =}{} +\sin\rho\cos\alpha\sin\theta\cos\phi\dpx{\alpha}
+\frac{\sin\rho\cos\theta\cos\phi}{\sin\alpha}\dpx{\theta}
\frac{\sin\rho\sin\phi}{\sin\alpha\sin\theta}\dpx{\phi}\right),\\
M_{02} =-i\left(\cos\rho\sin\alpha\sin\theta\sin\phi\dpx{\rho}
 \right.\\
 \left.\phantom{M_{02} =}{} +\sin\rho\cos\alpha\sin\theta\sin\phi\dpx{\alpha}
+\frac{\sin\rho\cos\theta\sin\phi}{\sin\alpha}\dpx{\theta}
\frac{\sin\rho\cos\phi}{\sin\alpha\sin\theta}\dpx{\phi}\right),\\
M_{03} =-i\left(\cos\rho\sin\alpha\cos\theta\dpx{\rho}
+\sin\rho\cos\alpha\cos\theta\dpx{\alpha}
-\frac{\sin\rho\sin\theta}{\sin\alpha}\dpx{\theta}\right),\\
M_{04} =-i\left(\cos\rho\cos\alpha\dpx{\rho}
-\sin\rho\sin\alpha\dpx{\alpha}\right).
\end{gather*}

The $O(1,4)$-invariant measure on $M_H$ is
\begin{gather*}
 d\mu= \sqrt{-g}\,dx^0dx^1 dx^2 dx^3=(\cos\rho)^{-4}  d\rho\, du,
\end{gather*}
where $du$ is the
$O(4)$-invariant measure on $S^3$.

\subsection{de Sitter group}
The universal covering of the de Sitter group is the symplectic
$Sp(2,2)$ group, which  is needed when dealing with half-integer spins.
It is suitably described as a subgroup of the group of $2\times2$
matrices with quaternionic coef\/f\/icients:
\begin{gather}
\label{sp}
Sp(2,2) = \left\{ g=
\begin{pmatrix}
a & b \\
c & d \end{pmatrix}
; \ a, b, c, d \in \H , \
 g^{\dagger}\gamma^0 g = \gamma^0 \equiv \begin{pmatrix}
   1   &  0  \\
    0  &  -1
\end{pmatrix}\right\}.
\end{gather}
We recall that the quaternion f\/ield as a multiplicative group is  $\H \simeq \R_{+}\times SU(2)$.
We write  the canonical basis for $\H \simeq \R^4$ as $(1\equiv e_4, e_i$ ($\simeq (-1)^{i+1}\sigma _{i})$ (in $2\times 2$-matrix notations), with $i=1, 2,3$: any   quaternion decomposes as
$q=(q^4, \vec{q})$ (resp.\ $q^{a} e_a, a=1,2,3,4$) in scalar-vector notations (resp.\ in Euclidean metric notation). We also recall that the multiplication law expli\-cit\-ly reads  in scalar-vector notation: $qq' = (q^4 {q'}^4 - \vec{q}\cdot \vec{q'}, {q'}^4\vec{q} + {q}^4\vec{q'} + \vec{q}\times \vec{q'})$. The (quaternionic) conjugate of $q=(q^4, \vec{q})$ is $\bar{q}=(q^4, -\vec{q})$, the squared norm is $\Vert q \Vert^2 = q\bar{q}$, and the inverse of a nonzero quaternion is $q^{-1} = \bar{q}/ \Vert q \Vert^2$. In (\ref{sp}) we have written  $g^{\dagger}= \bar{g}^{ t}$ for  the quaternionic
conjugate and transpose of the matrix~$g$.

Note that the def\/inition (\ref{sp}) implies the following relations between the matrix elements of $g \in Sp(2,2)$:
\begin{gather*}
\nonumber g = \begin{pmatrix}
  a    &   b \\
 c     &  d
\end{pmatrix} \in Sp(2,2)\\
\Leftrightarrow \ g^{-1} = \begin{pmatrix}
  \overline{a}    &   - \overline{c} \\
- \overline{b}     &  \overline{d}
\end{pmatrix}\Leftrightarrow \ \left\lbrace \begin{array}{ll}
    \Vert a  \Vert = \Vert d \Vert , &   \Vert b  \Vert = \Vert c \Vert  ,\\
     \Vert a  \Vert^2 - \Vert b \Vert^2 = 1 ,   &   \overline{a}b = \overline{c}d \Leftrightarrow a\overline{c} = b\overline{d} ,
\end{array} \right.
\end{gather*}
and we also note that $ \Vert a  \Vert =  \Vert d  \Vert  >  \Vert b  \Vert =  \Vert c  \Vert $, and ${\det}_{ 4\times4}g = 1$ for all $g\in Sp(2,2)$ when the latter are viewed as  $4\times 4$ complex matrices since $\H \simeq\R_+\times SU(2)$.

\subsection{1+3 de Sitter Clif\/ford algebra}

 The matrix
 \begin{gather*}
\gamma^0= \left(
\begin{array}{cc}
1 & 0 \\
0 & -1
\end{array} \right)
\end{gather*}
is part of the Clif\/ford algebra def\/ined by $\gamma^{\alpha}\gamma^{\beta} + \gamma^{\beta}
\gamma^{\alpha} = 2 \eta^{\alpha \beta}\ID$, the four other matrices having the
following form in this quaternionic representation:
\begin{gather*}
 \gamma^4= \left(
\begin{array}{cc}
0 & 1 \\
-1 & 0 \\
\end{array} \right) , \qquad
\gamma^k= \left(
\begin{array}{cc}
0 & e_k \\
e_k & 0
\end{array} \right), \qquad k= 1, 2, 3.
\end{gather*}
These matrices allow the following correspondence between points of~$\M_5$, or of the hyperboloid~$M_H$, and $2\times 2$ quaternionic matrices of the
form below:
\begin{gather*}
\M_5\ \mbox{or}\ M_H \ni x \lga \barra{x} \equiv x^{\al}\ga_{\al} = \left( \begin{array}{cc}
x^0 & - {\cal P} \\
\overline{\cal P}& -x^0 \\
\end{array} \right) \longleftrightarrow
	\mathcal{X} \equiv \begin{pmatrix}
 x^0     & {\cal P}   \\
   \overline{\cal P}   &  x^0
\end{pmatrix} = \barra{x}\gamma^0,
\end{gather*}
where ${\cal P} \equiv (x^4, \vec{x}) \in \H$. Note that we
have
\begin{gather*}
	x\cdot x= \barra{x} ^{\dag}\gamma^{0}\barra{x}\gamma ^{0} ,\qquad
	x^{\alpha} =\tfrac{1}{4}\mbox{tr}( \gamma^{\alpha}\barra{x}),\nonumber\\
	x\cdot x' = \tfrac{1}{4}\mbox{tr}(\barra{x}\barra{x'}), \qquad
	{\det}_{4\times4}\barra{x} = {\det}_{4\times4}\mathcal{X} = (x\cdot x)^2  .
\end{gather*}

\subsection{1+3 de Sitter group action}

Let $\Lambda  \in  SO_{0}(1,4)$ transform a wave plane as:
\begin{gather}\label{wpldstr}
	(Hx\cdot \xi )^{\sigma} \to
	(H(\Lambda^{-1}.x)\cdot\xi )^{\sigma} = (Hx\cdot \Lambda.\xi )^{\sigma}.
\end{gather}

Now, the action of $SO_{0}(1,4)$ on 4+1 Minkowski $\M_5$ amounts to the following $Sp(2,2)$ action on the de Sitter manifold or on the positive or negative null cone $\mathcal{C}^{\pm} = \{ \xi \in   \mathcal{C} \mid   \xi_0 \gtrless 0\}$.
\begin{gather}\label{dsact}
	Sp(2,2) \ni g :  \ \barra{x} \mapsto \barra{x}' = g\barra{x} g^{-1} \
	\Leftrightarrow   \ \mathcal{X}' =  g\mathcal{X}g^{\dag},
\end{gather}
and this precisely realizes the isomorphism $SO_{0}(1,4)
\to Sp(2,2)/\Z_2 $ through
\begin{gather*}
	SO_{0}(1,4) \ni \Lambda(g) : \   x \mapsto x' = \Lambda(g).x, \qquad
	\Lambda^{\alpha}_{\beta} = \tfrac{1}{4} \mbox{tr}\big(\gamma^{\alpha}g\gamma_{\beta}g^{-1}\big).
\end{gather*}

	Suppose $\det{\mathcal{X}} = 0$, i.e.\ $x \equiv \xi \in \mathcal{C}$.  To any $\xi = (\xi^0, \vec{\xi}, \xi^4) \equiv (\xi^0, \mathcal{P}) \in  \mathcal{C}$ there corresponds $v = v(\xi) \in S^3$ through:
\begin{gather*}
v = \frac{\mathcal{P}}{\xi^0} .
\end{gather*}
	Then the action \eqref{dsact} amounts to the following projective (or Euclidean conformal) action on the sphere $S^3$
\begin{gather*}
	SO_0(1,4) \ni \Lambda   : \ \xi \mapsto \xi' = \Lambda.\xi
	\nonumber\\
  \Leftrightarrow \  Sp(2,2) \ni g    : \  \left\{
	\begin{array}{l}
		\xi^0 \mapsto {\xi^0}' = \xi^0 \lVert c v + d\rVert^2, \\
		v \mapsto v' =  (a v + b)(cv + d)^{-1} \equiv g.v,
	\end{array}\right.
\end{gather*}
and $\mathrm{sgn}(\xi'^0) = \mathrm{sgn}(\xi^0)$.

Let us def\/ine matrix elements $\mathrm{T}^{-\sigma}_{L'l'm',Llm}$ (resp.\ $\mathcal{T}^{-\sigma}_{L'l'm',Llm}$) of the scalar (at least for the principal and complementary series) representations of $SO_0(1,4)$ (resp.\ $Sp(2,2)$) by
\begin{gather}\label{repso}
 	\Psi_{Llm}^{\sigma} (\Lambda^{-1}.x)
		 = \sum_{L'l'm'}\mathrm{T}^{-\sigma}_{L'l'm',Llm}(\Lambda)
		\Psi_{L'l'm'}^{\sigma} (x) ,  \\
 	\label{repsp}
 	\lVert c v + d\rVert^{2 \sigma}   \mathrm{Y}_{Llm}(g^{-1}.v)
		 =\sum_{L'l'm'}\mathcal{T}^{-\sigma}_{L'l'm',Llm}(g)   \mathrm{Y}_{L'l'm'}(v)  , \qquad g^{-1} =\begin{pmatrix}
   a   & b   \\
   c   &  d
\end{pmatrix}  .
\end{gather}
One can determine rather easily  the matrix elements (\ref{repsp}), and consequently  the elements~(\ref{repso}) through the two expansions of (\ref{wpldstr}).

\section{The massless minimally coupled f\/ield\\
 as an illustration of a Krein structure}\label{krein}

\subsection[The ''zero-mode'' problem]{The ``zero-mode'' problem}\label{zeromod}

We now turn our attention to the f\/irst element of the discrete series, which corresponds to  $\sigma=-3$, namely
 the massless minimally coupled f\/ield case. For $L\neq0$, we obtain the normalized modes
$\Psi_{-3;Llm}$ that we write  $\Psi_{Llm}$ for simplicity:
\begin{gather*}
\Psi_{Llm}(x)=\chi_{L}(\rho)Y_{Llm}(\Omega),
\end{gather*}
with
\begin{gather*}\chi_{L}(\rho)=\frac{H}{2}[2(L+2)(L+1)L]^{-\frac{1}{2}}
  \big(L e^{-i(L+2)\rho}+(L+2)e^{-iL\rho}\big).
\end{gather*}
As was already noticed,  the normalization
factor breaks down at $L=0$. This is the famous ``zero-mode'' problem, examined by many authors \cite{allen,allenfol,gareta}. In particular, Allen has shown that this zero-mode problem is responsible for the absence of a de Sitter invariant vacuum state for the mmc quantized  f\/ield. The non-existence, in the usual \textit{Hilbert} space quantization, of a de Sitter invariant vacuum state for the massless minimally coupled scalar f\/ield was at the heart of the  motivations of \cite{gareta}. Indeed, in order to circumvent this obstruction,  a Gupta--Bleuler type construction based on a Krein space structure was presented in \cite{gareta} for the quantization of the mmc f\/ield. One of the major advantages of this construction is the existence of a de Sitter invariant vacuum state. This is however not a Hilbert space quantization, in accordance with Allen's results. The rationale supporting the Krein quantization stems from de Sitter invariance requirements as is explained in the sequel.
The space generated by the $\Psi_{Llm}$ for $L\neq0$ is not a complete set of modes.
Moreover this set is not invariant under the action of the de~Sitter group.
Actually, an explicit computation gives
\begin{gather}
\label{ncov}
(M_{03}+iM_{04})\Psi_{1,0,0}=-i\frac{4}{\sqrt{6}}\Psi_{2,1,0}+\Psi_{2,0,0}+
\frac{3H}{4\pi\sqrt{6}},
\end{gather}
and the invariance is broken due to the last term. As a consequence,  canonical
f\/ield quantization applied to this set of modes yields a non covariant f\/ield,
and this  is due to  the appearance  of the  last term in (\ref{ncov}).
Constant functions are of course solutions to the f\/ield equation.
So one is led to deal with the space generated by the $\Psi_{Llm}$'s
and by a constant function denoted here by $\Psi_g$, this is
interpreted as a
``gauge'' state.
This space, which {\it is invariant under the de~Sitter group}, is the space of
physical states. However, as an inner-product space
equipped with the Klein--Gordon inner product, it is a degenerate space
because the state $\Psi_g$ is orthogonal to the whole space including
itself. Due to this degeneracy, canonical quantization applied to this set
of modes yields  a non covariant f\/ield (see~\cite{dbr} for a
detailed discussion of this fact).

Now, for $L=0$, as expected from equations (\ref{regsol}) and (\ref{singsol111}), the equation (\ref{eqrad}) is easily solved.
We obtain two independent solutions of the f\/ield equation, including the constant
function discussed above:
\begin{gather*}
\Psi_g=\frac{H}{2\pi}\qquad \mbox{and} \qquad
\Psi_s=-i\frac{H}{2\pi}\left\lbrack\rho+\frac{1}{2}\sin 2\rho \right\rbrack.
\end{gather*}
These two states are null norm. The  constant factors have been chosen in order to have
$\le\Psi_g,\Psi_s\re=1$.
We then  def\/ine $\Psi_{000}=\Psi_g+\Psi_s/2$. This is the ``true zero mode''
of Allen. We write  $\Psi_{000}=\Psi_{0}$ in the following.
With this mode, one obtains a complete set of strictly positive norm
modes $\Psi_{Lml}$ for $L\geq0$, but the space generated by these modes {\it is not
de~Sitter invariant}. For instance, we have
\begin{gather}
(M_{03}+iM_{04})\Psi_{0}=(M_{03}+iM_{04})\Psi_s\nonumber\\
\phantom{(M_{03}+iM_{04})\Psi_{0}}{}
=-i\tfrac{\sqrt{6}}{4}\Psi_{1,0,0}+-i\tfrac{\sqrt{6}}{4}\Psi_{1,0,0}^{*}
-\tfrac{\sqrt{6}}{4}\Psi_{1,1,0}-\tfrac{\sqrt{6}}{4}\Psi_{1,0,0}^{*}.\label{ncovd}
\end{gather}

Note the
appearance of negative norm modes in (\ref{ncovd}): this is the price
to pay in order to obtain
a fully covariant theory. The existence of these non physical states has led
 authors of \cite{gareta} to adopt
what they also  called \emph{Gupta--Bleuler field quantization}.
One of the essential ingredient of their procedure is the non vanishing inner products between  $\Psi_g$, $\Psi_s$ on one hand and
$\Psi_{Lml}$ and
$(\Psi_{Llm})^*$ for $L>0$ on the other hand:
\begin{gather*}
\le\Psi_{Llm},\Psi_{Llm}\re=1,\qquad
\le\Psi_{Llm}^*,\Psi_{Llm}^*\re=-1,\qquad
L>0 \qquad \mbox{and} \qquad
\le\Psi_s,\Psi_g\re=1.
\end{gather*}

\subsection[Gupta-Bleuler triplet and the mmc Krein structure]{Gupta--Bleuler triplet and the mmc Krein structure}\label{GBkrein}

In order to simplify the previous notations, let $K$ be the set of indices for
the positive norm modes, excluding the zero mode:
\begin{gather*}
K=\{(L,l,m)\in\N\times\N\times\Z;\  L\neq0,\  0\leq l\leq L,\
-l\leq
     m\leq l\},
\end{gather*}
     and let $K'$ be the same set including the zero mode:
     \begin{gather*}
K'=K\cup\{0\}.
\end{gather*}
As illustrated by (\ref{ncov}), the set spanned by the $\Psi_k$, $k\in K$
is not invariant under the action of the de Sitter group. On the other hand, we
obtain an invariant space by adding $\Psi_g$. More precisely, let us introduce
the space,
\begin{gather*}
\mathcal{K}=\Bigg\{c_g\Psi_g+\sum_{k\in K}c_k\Psi_k;\  c_g,c_k\in\C,\  \sum_{k\in
       K}|c_k|^2<\infty\Bigg\}.
\end{gather*}
Equipped with the Klein--Gordon-like inner product (\ref{kgie}), $\K$ is a
degenerate inner product space because the above orthogonal basis satisf\/ies to
\begin{gather*}
\le\Psi_k,\Psi_{k'}\re=\delta_{kk'}\quad  \forall\, k,k'\in K,\qquad
 \le\Psi_k,\Psi_g\re=0\quad  \forall \, k\in K, \qquad \mbox{and} \qquad \le\Psi_g,\Psi_g\re=0.
 \end{gather*}
It can be proved by conjugating the action (\ref{ncov}) under the $SO(4)$ subgroup
that $\K$ is invariant under the natural action of the de Sitter group. As a
consequence, $\K$ carries a unitary representation of the de Sitter group,  this
representation is indecomposable but not irreducible, and the null-norm subspace
$\n=\C\Psi_g$ is an uncomplemented invariant subspace.

Let us recall that the Lagrangian
\[
{\cal L}=\sqrt{|g|}g^{\mu\nu}\partial_\mu\Psi\partial_\nu \Psi
\]
of the free minimally coupled f\/ield is invariant when adding to $\Psi$ a
constant function. As a consequence, in the ``one-particle sector'' of the
f\/ield, the space of ``global gauge states'' is simply the  invariant
one dimensional
subspace   $\n=\C\Psi_g$. In the following, the space $\K$  is called
the (one-particle) physical space, but {\it stricto sensu} physical states are
def\/ined up to a constant and the space of physical states is $\K/\n$. The
latter is a Hilbert space carrying the unitary irreducible representation of the
de Sitter group $\Pi_{1,0}$.

If one attempts to apply the  canonical
quantization starting from a degenerate space of solutions, then one inevitably
breaks the covariance of the f\/ield~\cite{dbr}. Hence we must build a~non
degenerate invariant space of solutions $\HH$ admitting $\K$ as an invariant
subspace. Together with $\n$, the latter are constituent of the so-called Gupta--Bleuler triplet
$\n\subset\K\subset\HH$. The construction of $\HH$ is worked out as follows.

We f\/irst remark that the modes $\Psi_k$ and $\Psi_g$ do not form a
complete set of modes. Indeed, the  solution $\Psi_s$  does not belong to
$\K$ nor $\K+\K^{*}$ (where $\K^{*}$ is
the set of complex conjugates of~$\K$): in this sense, it is  not a
superposition of the modes $\Psi_k$ and $\Psi_g$. One way to prove this is to
note that $\le\Psi_s,\Psi_g\re=1\neq0$.

So we need a complete, non-degenerate and invariant inner-product space
containing $\K$ as a closed subspace. The smallest one fulf\/illing
these conditions is the following. Let $\HH_{+}$ be the Hilbert space
spanned by the modes $\Psi_{k}$ together with the zero-mode
$\Psi_{0}$:
\[
\HH_+=\Bigg\{c_{0}\phi_{0}+\sum_{k\in K}c_k\phi_k;\   \sum_{k\in
K}|c_k|^2<\infty\Bigg\}.
\]
We now def\/ine the total space $\HH$
by
\[
\HH=\HH_{+}\oplus\HH_{+}^{*},
\]
which {\it is} invariant, and we denote by $U$ the natural representation of the
de~Sitter group on $\HH$ def\/ined by: $U_g\Psi(x)=\Psi(g^{-1}x)$.
The space $\HH$ is def\/ined as a direct sum of an Hilbert space and an anti-Hilbert space (a
space with def\/inite negative inner product) which proves that $\HH$ is a
Krein space. Note that neither $\HH_+$ nor $\HH_+^{*}$ carry a representation of
the de Sitter group, so that the previous decomposition is not
covariant, although it is $O(4)$-covariant.
The following family is a pseudo-orthonormal basis for this Krein
space:
\begin{gather*}
\Psi_k, \ \ \Psi_k^{*} \  \ (k\in K),\ \ \Psi_0, \ \ \Psi_0^{*},
\end{gather*}
for which the non-vanishing inner products are
\begin{gather*}
\le\Psi_k,\Psi_k\re=\le\Psi_0,\Psi_0\re=1\qquad \mbox{and} \qquad
\le\Psi_k^{*},\Psi_k^{*}\re=\le\Psi_0^{*},\Psi_0^{*}\re=-1.
\end{gather*}

Let us once more insist on the presence of non physical states in
$\HH$. Some of them have  negative norm, but,
for instance, $\Psi_0$ is not a physical state ($\Psi_0\not\in\K$) in spite of
the fact that $\le\Psi_0,\Psi_0\re>0$: the condition of positivity of the inner
product is not a suf\/f\/icient condition for selecting physical states. Moreover
some non physical states go to negative frequency states when the
curvature tends to 0. Nevertheless mean values of observables are computed on
physical states and no  negative energy  appears.

We end this section by some general comments on the use of Krein space structures in quantum f\/ield theory. Already at the f\/lat space-time level, the usual Gupta--Bleuler treatment of gauge theories can be put into a Krein space setting \cite{Stro1}. For the two-dimensional massless scalar f\/ield in f\/lat spacetime, which is a model that mimics some of the features of gauge theories, Strocchi et al.\ have also followed in~\cite{Stro2} a Krein space approach. Indeed Krein structures can be viewed as a unif\/ied framework to treat gauge~-- and gauge-like~-- quantum f\/ield theories. Since a massless scalar f\/ield theory possesses a gauge-like invariance under the addition of a constant f\/ield, $\phi \rightarrow \phi+ \textrm{const}$,  our use of Krein structures to treat the ``massless'' scalar f\/ield on de Sitter space-time can be viewed as belonging to the same type of ideas used in the  works cited above. Nevertheless,  we apply the Krein space setting in a very dif\/ferent manner and the analogy ends here.

\section{Outline of future work}\label{outlinekrein}
The example of the mmc case is limpid: behind the Gupta--Bleuler and Krein structures lies the undecomposable nature of the involved de Sitter representation, the unitary irreducible $\Pi_{1,0}$ being realized on the coset $\HH/\mathcal{N}$, the one-dimensional null-norm space $\mathcal{N}$ being trivially cancelled under the action of the dif\/ferential operator $\partial_{\rho}$. This is the mark of an interesting cohomology accompanying the simple Lie group $Sp(2,2)$ \cite{pincsimon}. It is naturally appealing to approach  the mmc case with this algebraic point of view, and above all, to extend our analysis to all the elements of the scalar discrete series, with the hope that the obtained results will allow a complete covariant quantization of the corresponding f\/ields.

Another interesting direction for future research is the question of determining localization properties for these f\/ield theories through the use of modular localization techniques. In fact it seems that the ideas developed in \cite{BrGuLo}, render possible an ef\/fective  exploration of the localization properties of the de Sitterian f\/ields carrying the discrete series.

\appendix

\section[Lie algebra $B_2$: a minimal glossary]{Lie algebra $\boldsymbol{B_2}$: a minimal glossary}\label{B2}

\paragraph{Def\/inition.}
\begin{gather*}
sp(2,2)=\left\{\left. \mat{\vec{v} & p \\ \tilde{p} & \vec{w}} \right| (0,\vec{v}),(0,\vec{w}),p\in \HC_{\C}  \right\}.
\end{gather*}

\paragraph{Operations in $\boldsymbol{\HC_{\C}}$:}
\begin{itemize}\itemsep=0pt
\item $p,q\in \HC_{\C}$, $pq:=(p_0q_0-\vec{p}\vec{q},p_0 \vec{q}+q_0\vec{p}+\vec{p}\times\vec{q})$,
\item $z\in \HC_{\C}$, $\tilde{z}:=(z_0,-\vec{z})$,
\item $z\in \HC_{\C}$, $\|z\|^2:=z\tilde{z}=z_0^2+z_1^2+z_2^2+z_3^2$,
\item $z\in \HC_{\C}$, $z\neq0$, $z^{-1}=\frac{1}{\|z\|^2} \tilde{z}$.
\end{itemize}

\paragraph{Cartan subalgebra
 $\boldsymbol{\mathfrak{h}=\spa \{H_1,H_2\}}$.}
\begin{gather*}
H_1=\frac{1}{2}\mat{0&1\\1&0}, \qquad H_2=\frac{i}{2}\mat{e_3&0\\0&e_3},\\
[H_1,H_2]=0, \qquad  N(\mathfrak{h})=\mathfrak{h},\\
\left[H_1,X_{-1-1}\right]=-X_{-1-1},\qquad  \left[H_2,X_{-1-1}\right]=- X_{-1-1},\\
X_{-1-1}=\frac{1}{4}\mat{e_1-i e_2 & e_1-i e_2 \\ -e_1+i e_2 & -e_1+i e_2}, \\
\left[H_1,X_{-10}\right]  = -X_{-10}, \qquad \left[H_2,X_{-10}\right]  = 0,       \qquad
 X_{-10}  = \frac{1}{2}\mat{e_3 & e_3 \\ -e_3 & -e_3}, \\
\left[H_1,X_{-11}\right]  = -X_{-11}, \qquad \left[H_2,X_{-11}\right]  = X_{-11},	  \qquad
 X_{-11} =\frac{1}{4}\mat{e_1+i e_2 & e_1+i e_2 \\ -e_1-i e_2 & -e_1-i e_2},\\
\left[H_1,X_{0-1}\right] =0       , \qquad  \left[H_2,X_{0-1}\right] =-X_{0-1},  \qquad
 X_{0-1} =\frac{1}{2}\mat{ e_1-i e_2 & 0 \\ 0 & e_1 -i e_2}, \\
\left[H_1,X_{01}\right]  =0       , \qquad \left[H_2,X_{01}\right]  = X_{01},   \qquad
 X_{01}  =\frac{1}{2}\mat{ e_1+i e_2 & 0 \\ 0 & e_1+i e_2}, \\
\left[H_1,X_{1-1}\right] =X_{1-1} , \qquad \left[H_2,X_{1-1}\right] =- X_{1-1}, \qquad
 X_{1-1} =\frac{1}{4}\mat{e_1-i e_2 & -e_1+i e_2 \\ e_1-i e_2 & -e_1+i e_2}, \\
\left[H_1,X_{10}\right]  =X_{10}  , \qquad \left[H_2,X_{10}\right]  =0,     \qquad
 X_{10}  =\frac{1}{2}\mat{e_3 & -e_3\\ e_3 & -e_3}, \\
\left[H_1,X_{11}\right]  =X_{11}  , \qquad \left[H_2,X_{11}\right]  =X_{11},	\qquad
 X_{11}   = \frac{1}{4}\mat{e_1+i e_2 &-e_1-i e_2 \\ e_1+i e_2 & -e_1-i e_2}.
\end{gather*}

\paragraph{Root system $\boldsymbol{R}$.}
\begin{gather*}
\alpha_{ij}(H_1)=i, \qquad \alpha_{ij}(H_2)=j,\\
\alpha_{-1-1}, \ \ \alpha_{-10}, \ \ \alpha_{-11}, \ \ \alpha_{0-1},
\ \ \alpha_{01}, \ \ \alpha_{1-1}, \ \ \alpha_{10}, \ \ \alpha_{11}.
\end{gather*}

\paragraph{Basis:}
 $B=(\beta_1,\beta_2)=(\alpha_{10},\alpha_{-1,1})$.

\paragraph{Simple roots:} $\beta_1$, $\beta_2$.

\paragraph{Roots in the basis $\boldsymbol{B}$:}
$(\beta_1)_B=(1,0)$, $(\beta_2)_B=(0,1)$, $(\alpha_{0,1})_B=(1,1)$, $(\alpha_{1,1})_B=(2,1)$,
$(\alpha_{-1,0})_B=(-1,0)$,  $(\alpha_{1,-1})_B=(0,-1)$, $(\alpha_{0,-1})_B=(-1,-1)$, $(\alpha_{-1,-1})_B=(-2,-1) $.

\paragraph{Positive roots:}
$\alpha_{-1,1}$, $\alpha_{1,0}$, $\alpha_{0,1}$, $\alpha_{1,1}$.

\paragraph{Highest root:}  $\Theta=2\beta_1+\beta_2$.
\begin{gather*}
\left[H,X_{\alpha}\right] = \alpha(H)X_{\alpha},\\
\left[X_{\alpha},X_{\beta}\right] = 0 \qquad {\rm if} \ \ \alpha+\beta\notin R\cup\{0\},\\
\left[X_{\alpha},X_{-\alpha}\right] = -H_{\alpha}, \\
\left[X_{\alpha},X_{\beta}\right] = N_{\alpha\beta}X_{\alpha+\beta} \qquad {\rm if} \ \ \alpha+\beta\in R, \\
\alpha(H_{\alpha}) = 2,
\\
H_{01}=2H_2=i\mat{e_3&0\\0&e_3}, \qquad H_{0-1} = -2H_2=-i\mat{e_3&0\\0&e_3},\\
H_{10} = 2H_1=\mat{0&1\\1&0},\qquad H_{-10} = -2H_1=-\mat{0&1\\1&0}  ,\\
H_{11} = H_1+H_2=\frac{1}{2}\mat{i e_3&1\\1&ie_3},\qquad H_{-1-1} = -H_1-H_2=-\frac{1}{2}\mat{i e_3&1\\1&ie_3},\\
H_{1-1} = H_1-H_2=\frac{1}{2}\mat{-i e_3&1\\1&-ie_3},\qquad H_{-11} = -H_1+H_2=\frac{1}{2}\mat{i e_3&-1\\-1&ie_3}.
\end{gather*}

\paragraph{Killing form  $\boldsymbol{K(x,y)}$ and duality.}
 Restricted to the Cartan subalgebra $\mathfrak{h}$
\begin{gather*}
x=a_1 H_1 + a_2 H_2, \qquad y=b_1 H_1 + b_2 H_2, \qquad a_{1,2}, b_{1,2} \in \C,\\
K(x,y)|_{\mathfrak{h} \times \mathfrak{h}}=\sum_{\alpha \in R}{\alpha(x)\alpha(y)}=6 a_1 b_1 + 6 a_2 b_2.
\end{gather*}

Dual elements of $\gh$ corresponding to roots
\begin{gather*}
\forall \, x\in B_2 \qquad \alpha(x)=K(h_{\alpha},x),
\\
h_{-1-1} = \tfrac{1}{6} H_{-1-1},\qquad h_{-10} = \tfrac{1}{12} H_{-10}, \qquad h_{-11} = \tfrac{1}{6} H_{-11},
\qquad  h_{0-1} = \tfrac{1}{12} H_{0-1},  \\
h_{11} = \tfrac{1}{6} H_{11},\qquad  h_{10} = \tfrac{1}{12} H_{10}, \qquad h_{1-1} = \tfrac{1}{6} H_{1-1},\qquad h_{0-1} = \tfrac{1}{12} H_{0-1}.
\end{gather*}

\paragraph{Root geometry.}

\paragraph{Length of simple roots, angle between them:}
\begin{gather*}
\<\alpha,\beta\>:=K(h_{\alpha},h_{\beta}), \qquad \alpha,\beta \in R,\\
 \<\beta_1,\beta_1\>=\tfrac{1}{6}, \qquad \<\beta_2,\beta_2\>=\tfrac{1}{3}, \qquad \<\beta_1,\beta_2\>=-\tfrac{1}{6},\\
 \cos \vartheta=\frac{\<\beta_1,\beta_2\>}{\sqrt{\<\beta_1,\beta_1\>}\sqrt{\<\beta_2,\beta_2\>}}=-\tfrac{\sqrt{2}}{2}  \ \Rightarrow \ \vartheta=\tfrac{3}{4}\pi.
\end{gather*}

\paragraph{Cartan matrix:}
\begin{gather*}
A=\mat{2 & -1 \\ -2 & 2}.
\end{gather*}

\paragraph{Fundamental weights:}
\begin{gather*}
\alpha^{\vee}:=\frac{2\alpha}{\<\alpha,\alpha\>}, \qquad \alpha \in  R,\qquad
 \<\beta_i^{\vee},\omega_j \>=\delta_{ij},\\
 \omega_1= \beta_1+\tfrac{1}{2}\beta_2, \qquad \omega_2= \beta_1+\beta_2.
\end{gather*}

\begin{figure}[th]
\centerline{\includegraphics{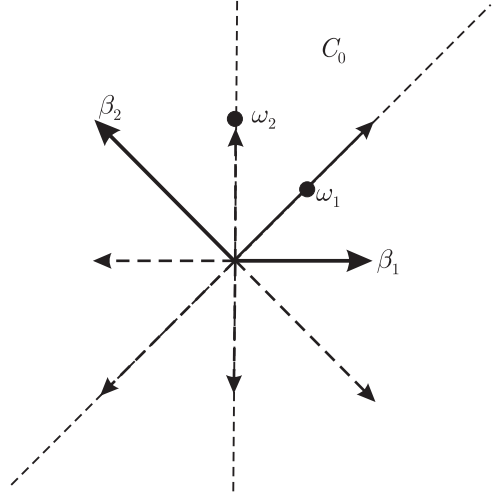}}
	\caption{Root system of $B_2$ and fundamental weights.}
	\label{fig:RootSystem}
\end{figure}

\paragraph{Root system diagram.} See Fig.~\ref{fig:RootSystem}.

\paragraph{Weyl orbit of a weight.}
$(n_1, n_2)$ are coef\/f\/icients in $\omega_1$, $\omega_2$ basis
\begin{gather*}
 (n_1, n_2) , \ \ (-n_1, n_1 + n_2) , \ \ (n_1 + 2 n_2, -n_2), \ \ (-n_1 - 2 n_2, n_1 + n_2),\\
 (n_1 + 2 n_2, -n_1 - n_2) , \ \ (-n_1 - 2 n_2, n_2), \  \ (n_1, -n_1 - n_2), \ \ (-n_1, -n_2).
\end{gather*}

\paragraph{Dimension of representation.}{\samepage
\paragraph{Weyl formula:}
\begin{gather*}
\dim  V = \prod_{\alpha\in R}{\frac{\<\lambda_0+\varrho,\alpha\>}{\<\varrho,\alpha \>}},
\end{gather*}
where} $\varrho=1/2\sum\limits_{\alpha\in R^+}\alpha=2\beta_1+3/2 \beta_2 $ and $\lambda_0$ is a dominant weight, i.e.\ the weight with non-negative integer coef\/f\/icients in the basis of fundamental weights $\omega_{1,2}$, $\lambda_0=(n_1+n_2)\beta_1+(n_1/2+n_2)\beta_2$, $n_{1,2}\in \N_0$.
\begin{gather*}
\dim V=\tfrac{1}{6}(n_1+1)(n_2+1)(n_1+2 n_2+3)(n_1+n_2+2).
\end{gather*}

\paragraph{Eigenvalues of Casimir:}
\begin{gather}
\label{casimirfin}
Q_{\lambda_0}(n_1,n_2)=-\tfrac{1}{2}\left( n_1^2+2n_2^2+2 n_1 n_2 + 4 n_1 +6 n_2\right) .
\end{gather}

\paragraph{Weyl equivalence with UIR.}
Two irreps are \emph{Weyl equivalent} if they share  same Casimir  eigenvalue. Comparing (\ref{casimirfin}) with (\ref{casim1}) gives the possible solutions:
\begin{gather}
\label{weyleq1}
  n_1 = -2q  , \qquad  n_2 = p+q -1    , \\
  \label{weyleq2}
    n_1 = 2q-2  , \qquad  n_2 = p-q    , \\
      n_1 = q-1  , \qquad  n_2 = -2p-2   ,   \nonumber\\
        n_1 = -2p-2    , \qquad  n_2 = p-q    .\nonumber
        \end{gather}
Only (\ref{weyleq1}) for $q=0$, $p\geq 1$,  and (\ref{weyleq2}) for $q \geq 1$, $p\geq 1$, yield possible Weyl equivalence between f\/inite-dimensional irreps $(n_1,n_2)$  and UIR's $\Pi_{p,q}^{\pm}$ in the discrete series.

\paragraph{Characters:}
\begin{gather*}
\xi=m_1 \omega_1+m_2 \omega_2,\\
\chi_{n_1,n_2}(m_1,m_2)=\frac{e^{\frac{1}{12} (-2 m_1 n_1-3 m_2 n_1-3 m_1 n_2-4 m_2 n_2)}}{\big(-1+e^{m_1/12}\big) \big(-1+e^{m_2/6}\big) \big(-1+e^{\frac{m_1+m_2}{6}}\big) \big(-1+e^{\frac{1}{12} (m_1+2 m_2)}\big)}  \\
\phantom{\chi_{n_1,n_2}(m_1,m_2)=}{}
\times\big(-e^{\frac{1}{12} ((m_1+2 m_2) (3+2 n_1)+4 m_1 n_2+6 m_2 n_2)} -e^{\frac{1}{12} (m_1+2 (m_1+m_2) (n_1+n_2))} \\
\phantom{\chi_{n_1,n_2}(m_1,m_2)=}{}
+e^{\frac{1}{12} (m_1+4 m_2+2 (m_1+2 m_2) (n_1+n_2))}+ e^{\frac{1}{12} (2 m_2 (n_1+n_2)+m_1 (n_1+2 n_2))}\\
\phantom{\chi_{n_1,n_2}(m_1,m_2)=}{}
+e^{\frac{1}{12} (3 m_1+2 m_2+2 (m_1+m_2) (n_1+2 n_2))}-e^{\frac{1}{12} (2 m_2+(m_1+2 m_2) (n_1+2 n_2))}\\
 \phantom{\chi_{n_1,n_2}(m_1,m_2)=}{}
+e^{\frac{1}{12} m_1 (4+3 n_1+4 n_2)}\big(-e^{\frac{1}{3} m_2 (1+n_1+n_2)}+e^{\frac{1}{6} m_2 (3+2 n_1+3 n_2)}\big)\big).
\end{gather*}

\subsection*{Acknowledgements}

P.~Siegl appreciates the support of CTU grant No.$\!$ CTU0910114 and MSMT project No.$\!$ LC06002.

\pdfbookmark[1]{References}{ref}
\LastPageEnding

\end{document}